\documentclass[apj]{emulateapj}

\usepackage{graphicx,times}
\usepackage{subfigure}
\newcommand{\be}{\begin{equation}}
\usepackage{threeparttable}
\usepackage{booktabs}
\newcommand{\ee}{\end{equation}}
\newcommand{\bea}{\begin{eqnarray}}
\newcommand{\eea}{\end{eqnarray}}

\usepackage{enumerate}
\usepackage{amsmath}
\usepackage{cases}
\usepackage{longtable}
\usepackage{hyperref}
\usepackage{epstopdf}
\usepackage{amsmath,bm}
\usepackage{amssymb}
\usepackage{natbib}
\usepackage{morefloats}
\usepackage{multirow}
\usepackage{array}
\usepackage{verbatim}

\begin{document}

\title{Turbulence in a self-gravitating molecular cloud core}

\author{Siyao Xu\altaffilmark{1,2} and Alex Lazarian\altaffilmark{1} }

\altaffiltext{1}{Department of Astronomy, University of Wisconsin, 475 North Charter Street, Madison, WI 53706, USA; 
sxu93@wisc.edu,
lazarian@astro.wisc.edu}
\altaffiltext{2}{Hubble Fellow}

\begin{abstract}

Externally driven interstellar turbulence plays an important role in shaping the density structure in molecular clouds. 
Here we study the dynamical role of internally driven turbulence in a self-gravitating molecular cloud core. 
Depending on the initial conditions and evolutionary stages, we find that a self-gravitating core in the presence of 
gravity-driven turbulence can undergo
constant, decelerated, and accelerated infall, and thus has various radial velocity profiles. 
In the gravity-dominated central region, a higher level of turbulence results in 
a lower infall velocity, a higher density, and a lower mass accretion rate. 
As an important implication of this study, 
efficient reconnection diffusion of magnetic fields 
against the gravitational drag naturally occurs due to the gravity-driven turbulence, without invoking externally driven turbulence.

\end{abstract}


\section{Introduction}

The interstellar medium (ISM) is turbulent 
(e.g., \cite{Armstrong95,CheL10}). 
The interstellar turbulence plays a significant role in physical processes including 
the star formation
\citep{Mckee_Ostriker2007,Fed12},
cosmic ray propagation
\citep{Sca04,XY13,Xu16,XLcr18},
dynamo amplification 
\citep{Bec96,Bran05,XL16,XLr17,XL17,XuG19}
and turbulent reconnection 
\citep{LV99,Laz14,LEC12}
of interstellar magnetic fields, 
and formation and evolution of interstellar density structure  
\citep{Pad01,Burk09,XuJ19},
accounting for observations, 
e.g., interstellar scattering of Galactic pulsars 
\citep{Cor85,Ric90,XuZ17},
rotation measure fluctuations 
\citep{MS96,Hav08,XuZ16},
fluctuations in synchrotron intensity and polarization 
\citep{Gae11,LP12,LP16}.

Supernova explosions are believed as a dominant source of turbulent energy on length scales of the order of $10-100$ pc
\citep{Kor99,Hav08}.
The injected turbulence cascades down toward smaller length scales 
\citep{Armstrong95,CheL10,Chep10,Qi18}. 
On small length scales in contracting dense cores in molecular clouds, 
an ``adiabatic heating" mechanism acts to amplify the internal turbulence due to compression 
\citep{RobG12}, 
where the gravitational potential energy is converted to the turbulent kinetic energy 
\citep{Sca82,Sur10}. 
The resulting additional internal turbulent pressure support is expected to affect the dynamics and evolution of collapsing cores, as well as their 
density and velocity profiles 
\citep{Lee15,Mu15}.
In this study, we incorporate the gravity-driven turbulence and investigate the dynamical evolution of a spherical self-gravitating core. 
The self-similar behavior of a collapsing sphere has been extensively studied both analytically and numerically
\citep{Lar69,Pen69,Sh77,Hun77,Fos93,Fat04,Lou04}. 
Here we follow the analytical approach of 
\cite{Sh77}
to solve the hydrodynamic equations,
but focus on the differences in solutions due to the presence of turbulence, 
which was not considered in the original formalism.

Apart from its importance in influencing the dynamics of molecular cloud cores, 
turbulence can also effectively enhance the diffusion efficiency of magnetic fields by 
enhancing their reconnection efficiency. 
In this work, we will also discuss the implication of gravity-driven turbulence on reconnection diffusion (RD) of magnetic fields. 
The paper is organized as follows. In \S 2, by solving the hydrodynamic equations involving the internal turbulent pressure, 
we analyze the dynamical effect of gravity-driven turbulence on the gravitational collapse of a spherical core. 
In \S 3, we discuss the implication on the RD of magnetic fields arising from the gravity-driven turbulence.  
The conclusions are provided in \S 4.

\section{Self-similar collapse of a self-gravitating turbulent sphere}

We consider a spherical geometry for a self-gravitating and isothermal sphere. 
The governing equations include the continuity equation in terms of mass $M$, 
the continuity equation in terms of density $\rho$,
and the momentum equation
\citep{Shu92}: 
\begin{subequations}\label{eq: setf}
\begin{align}
&  \frac{\partial M}{\partial t} +  u \frac{\partial M}{\partial r}= 0 ,  ~~  \frac{\partial M}{\partial r} = 4 \pi r^2 \rho, \label{eq:setf1}   \\ 
&   \frac{\partial \rho}{\partial t} + \frac{1}{r^2} \frac{\partial (r^2 \rho u)}{\partial r} = 0 , \\
&   \frac{\partial u}{\partial t} + u \frac{\partial u}{\partial r} + \frac{1}{\rho} \frac{\partial (\rho v_t^2 + \rho a^2)}{\partial r} + \frac{G M }{r^2} = 0 , \label{eq:setf3}
\end{align}
\end{subequations}
where $|u|$ is the fluid speed, $a$ is the sound speed, and $v_t$ is the turbulent speed. 
In the absence of external driving,
the turbulence in a contracting gas is amplified via the ``adiabatic heating" mechanism,
that is, 
turbulence adiabatically heats during contraction
\citep{RobG12}.
On the other hand, the turbulence dissipates as the turbulent energy cascades toward smaller scales. 
Under the effects of ``adiabatic heating" and dissipation of turbulence, $v_t$ follows 
\citep{RobG12,Mu15}:
\begin{equation}\label{eq: turad}
   \frac{\partial v_t}{\partial t} + u \frac{\partial v_t}{\partial r}  + (1 + \eta \frac{v_t}{ u}) \frac{v_t u}{r} =0,
\end{equation}
where the two terms in the brackets correspond to the turbulence driving and dissipation, respectively, 
and the parameter $\eta$ represents the efficiency of turbulent energy cascade. 
Both the turbulent motion driven by gravitational contraction and the thermal motion of gas 
contribute to the pressure support against gravity in Eq. \eqref{eq:setf3}.

To solve Eq. \eqref{eq: setf}, we follow the analytical approach presented in 
\citet{Sh77}
and combine the radius $r$ and the time $t$ into a dimensionless variable  
\begin{equation}
    x = \frac{r}{at}.
\end{equation}
We then look for a similarity solution of the form 
\begin{equation}\label{eq: ssef}
\begin{aligned}
  &   \rho(r,t) = \frac{\alpha(x)}{4 \pi G t^2},    
     ~~~~~M(r,t) = \frac{a^3 t}{G} m(x),   \\
  &   u(r,t) = a v(x),   
     ~~~~~~~v_t(r,t) = Ca v(x),
\end{aligned}
\end{equation}
where the dimensionless variables $\alpha$, $m$, and $v$ are the reduced density, mass, and fluid speed, 
and $G$ is the gravitational constant. 
Besides, $|C| \leq 1$ is the ratio of $v_t$ to $|u|$ and also the ratio of the turbulent eddy-turnover rate to the gravitational contraction rate, 
\begin{equation}\label{eq: synetc}
      |C| = \frac{v_t/r}{|u|/r}.
\end{equation}
For the contraction induced turbulence, 
both simulations and physical considerations 
suggest that $v_t$ tracks $|u|$ and tends to synchronize with $|u|$
\citep{RobG12,Mu15}.
\footnote{The synchronization is expected to be stable as the turbulent eddies are compressed on their turnover timescales
\citep{RobG12}.}
Therefore, we consider $C$ as a constant.

By substituting Eq. \eqref{eq: ssef} into Eqs. \eqref{eq: setf} and \eqref{eq: turad}, we find 
\begin{subequations}\label{eq: sscet}
\begin{align}
   &  ~~~~~~~~~~~~~~~~~~~~~~~~~~ m = x^2 \alpha (x-v), \label{eq: mnr} \\
   & \Big \{(x-v)\big[ (x-v) - 2 C^2 v\big] - (1+ C^2 v^2 )\Big\} \frac{dv}{dx}    \nonumber \\
   &  = (x-v) \Big[\alpha (x-v) - \frac{2}{x} (1+ C^2 v^2 )\Big],  \label{eq:cp1}  \\
  &  \Big\{(x-v)\big[(x-v)-2 C^2 v \big]  - (1+ C^2 v^2 )\Big\} \frac{d\alpha}{dx}    \nonumber \\
  &  = \alpha (x-v) \Big\{\alpha - \frac{2}{x} \big[(x-v) - 2 C^2 v \big] \Big\},    \label{eq:cp2}  \\
   &~~~~~~~~~~~~~~~~~~~ (x-v) \frac{dv}{dx} = (1 +  \eta C) \frac{v^2}{x} . \label{eq: c3}
\end{align}
\end{subequations}
The ratio of the gravitational force to the pressure gradient force is 
\begin{equation}\label{eq: genratr}
     \mathcal{R} = \frac{\frac{GM}{r^2}}{\frac{1}{\rho}\frac{\partial (\rho v_t^2 + \rho a^2)}{\partial r}}
     \sim \frac{   \alpha x (x-v)}{ C^2v^2 +  1},
\end{equation}
where the expressions in Eqs. \eqref{eq: ssef} and \eqref{eq: mnr} are used. 
If the effect of turbulence is negligible, it becomes 
\begin{equation}\label{eq: ratthe}
   \mathcal{R}_\text{the} 
    \sim \alpha x (x-v).
\end{equation}
For the ``inside-out" collapse of a singular isothermal sphere considered in 
\citet{Sh77}, 
the hydrodynamic signal propagates at the speed of sound. 
The envelope at $x>1$ can remain in the initial hydrostatic state, while the interior at $x<1$ undergoes gravitational infall. 
Here we incorporate the effect of self-driven turbulence. In the case of 
highly supersonic turbulence, i.e., $Cv \gg 1$,
$\mathcal{R}$ is approximately 
\begin{equation}\label{eq: ratur}
  \mathcal{R}_\text{tur} 
  \sim \frac{   \alpha x (x-v)}{  C^2  v^2},
\end{equation}
which is smaller than $\mathcal{R}_\text{the}$.

In various asymptotic limits, the solution to the coupled Eqs. \eqref{eq:cp1} and \eqref{eq:cp2} has different behaviors. 
We start with the limit $x \gg |v|$, i.e., $ r \gg |u| t  \geq v_t t$. 
It is beyond the radius where the hydrodynamic signals carried by turbulence can reach. 
We consider different cases with small and large initial infall velocities. 

Case (1): $x \rightarrow \infty$ ($t \rightarrow 0$), $v \rightarrow 0 $, $\alpha \ll 1$. 

At an initial state, if the infall velocity is sufficiently small, the effect of turbulence is negligible. 
This initial state can be treated as the
case of collapse of a singular isothermal sphere at a large $x$ considered in 
\citet{Sh77},
where turbulence was not taken into account. 
One can easily obtain the solution: 
\begin{subequations}\label{eq: afssvi}
\begin{align}  
     v &= - (A-2) x^{-1} ,     \\
     \alpha &= A x^{-2},   \label{eq: fslvden}
\end{align}
\end{subequations}
and 
\begin{equation}\label{eq: asrmsv}
   m \approx  \alpha x^3 = A x,
\end{equation}
where the constant $A$ should not be smaller than $2$ as $v$($\leq 0$) is an inward velocity. 


Case (2): $x \rightarrow \infty$ ($t \rightarrow 0$), $C v \gg 1$, $\alpha \ll 1$. 

This initial condition with a large infall velocity
allows the generation of supersonic turbulence. 
Accordingly, Eqs. \eqref{eq:cp1} and \eqref{eq:cp2} can be approximated by:
\begin{subequations}
\begin{align} 
       \frac{dv}{dx} 
   &  =  \alpha  - \frac{2 C^2 v^2}{x^2}  ,  \label{eq: lxlvc} \\
    x    \frac{d\alpha}{dx} 
   & = - 2\alpha , \label{eq: lxsd}
\end{align}
\end{subequations}
We find the solution 
\begin{subequations}\label{eq: csolv}
\begin{align}
     v 
     & \approx -v_2, \label{eq: clvv}\\
    \alpha 
     &= \alpha_2 x^{-2},
\end{align}
\end{subequations}
where $v_2$ and $\alpha_2$ are constants and 
$\alpha_2 \leq 2 C^2 v_2^2$.
With the same density profile as in Case (1), the asymptotic form of $m$ is the same as Eq. \eqref{eq: asrmsv},
\begin{equation}
      m \approx \alpha_2 x.
\end{equation}
To have a constant $v$, it requires $\eta  = -1/C$ (Eq. \eqref{eq: c3}). 
With comparable rates of turbulence driving and dissipation, the resulting turbulence has a constant $v_t$.
The above result shows that the system has undisturbed infall motions.

The different initial conditions in Case (1) and Case (2) lead to different behaviors of the subsequent collapse. 
We next consider the limit $x \ll |v|$, i.e., $ r \ll |u| t $, and rewrite 
Eqs. \eqref{eq:cp1} and \eqref{eq:cp2} as 
\begin{subequations}\label{eq: datvd}
\begin{align} 
 & \frac{dv}{dx} 
     =  \frac{1}{1 +  C^2} \Big(\alpha  + \frac{2C^2}{x}   v \Big), \label{eq:indyv} \\
 &     \frac{d\alpha}{dx} 
    = -\frac{\alpha }{(1+C^2 ) v} \Bigg[ \alpha + \frac{2 (1 + 2 C^2)}{x} v \Bigg].
\end{align}
\end{subequations}
The ratio of the two terms in the brackets in Eq. \eqref{eq:indyv} reflects the relative importance of gravitational contraction and 
turbulent pressure support (see Eq. \eqref{eq: ratur}). 
In the regime where the turbulent pressure 
plays a dominant role, i.e., 
$\mathcal{R}_\text{tur} \ll 1$, we can further simplify the above equations and have 
\begin{subequations}
\begin{align}  
 & \frac{dv}{dx} 
     =  \frac{2C^2}{1 +  C^2}    \frac{v}{x}    , \label{eq:vlvtd} \\
 &     \frac{d\alpha}{dx} 
    = -\frac{2 (1 + 2 C^2) }{1+C^2  }    \frac{\alpha}{x} ,
\end{align}
\end{subequations}
We derive the solutions as 
\begin{subequations}\label{eq: decasol}
\begin{align}
  &  v = -v_1 x^{\frac{2C^2}{1 +  C^2} },  \label{eq: vahn}  \\
  & \alpha = \alpha_1 x^{-\frac{2 (1 + 2 C^2) }{1+C^2  }  },
\end{align}
\end{subequations}
where $v_1$ and $\alpha_1$ are constants. 
As $|v|$ decreases with decreasing $x$, it shows that the turbulent pressure dominates the dynamics and causes deceleration of the infall. 
Given the above expressions, $m$ (Eq. \eqref{eq: mnr}) is approximately 
\begin{equation}\label{eq: asmbtp}
      m \approx \alpha_1 v_1 .
\end{equation}
By comparing Eq. \eqref{eq:vlvtd} with Eq. \eqref{eq: c3}, we find that the corresponding relation between $\eta$ and $C$ is
\begin{equation}\label{eq: etcred}
     \eta = - \frac{1+ 3 C^2}{C (1 + C^2)}.
\end{equation}
With $\eta |C| > 1$, the dissipation is more efficient than driving. 
Thus $v_t$ decreases with decreasing $x$.

In the 
regime where the gravitational contraction is more important than the turbulent support, i.e., 
$\mathcal{R}_\text{tur} \geq 1$,
the solution to Eq. \eqref{eq: datvd} is 
\begin{subequations}\label{eq: coffs}
\begin{align}
  &  v = - \frac{2 \alpha_0}{1+ 5 C^2} x^{-\frac{1}{2}}, \label{eq: ffves} \\
  &  \alpha = \alpha_0 x^{-\frac{3}{2}},
\end{align}
\end{subequations}
and 
\begin{equation}
      m \approx  \frac{2 \alpha_0^2}{1+ 5 C^2} , 
\end{equation}
where $\alpha_0$ is a constant. 
The above result has the same scaling as that of a free-fall collapse, 
as expected in the regime dominated by self-gravity.
Eq. \eqref{eq: ffves} indicates the relation
\begin{equation}\label{eq: trcrff}
    \eta \sim -\frac{1}{2C},
\end{equation}
showing enhanced turbulence toward smaller $x$ with $\eta |C| < 1$. 
The above relations between $C$ and $\eta$ in both regimes with decelerated and accelerated infall, 
i.e., Eq. \eqref{eq: etcred} with $C=-1$ and Eq. \eqref{eq: trcrff}, 
are consistent 
with earlier numerical simulations of contracting turbulence 
\citep{RobG12}.

\begin{figure*}[htbp]
\centering   
\subfigure[$A=2.001$]{
   \includegraphics[width=8.5cm]{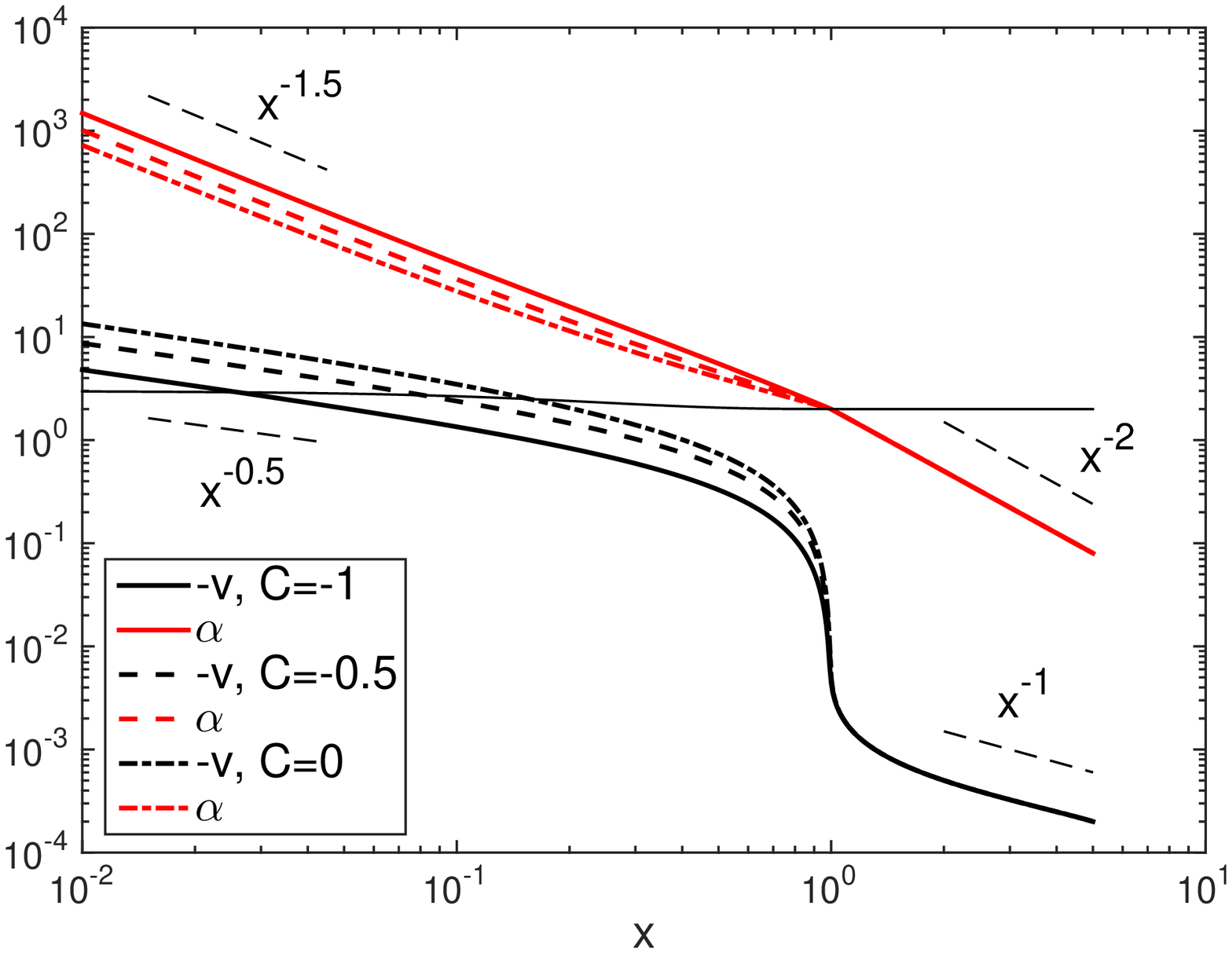}\label{fig: svda}}
\subfigure[$A=2.001$]{
   \includegraphics[width=8.5cm]{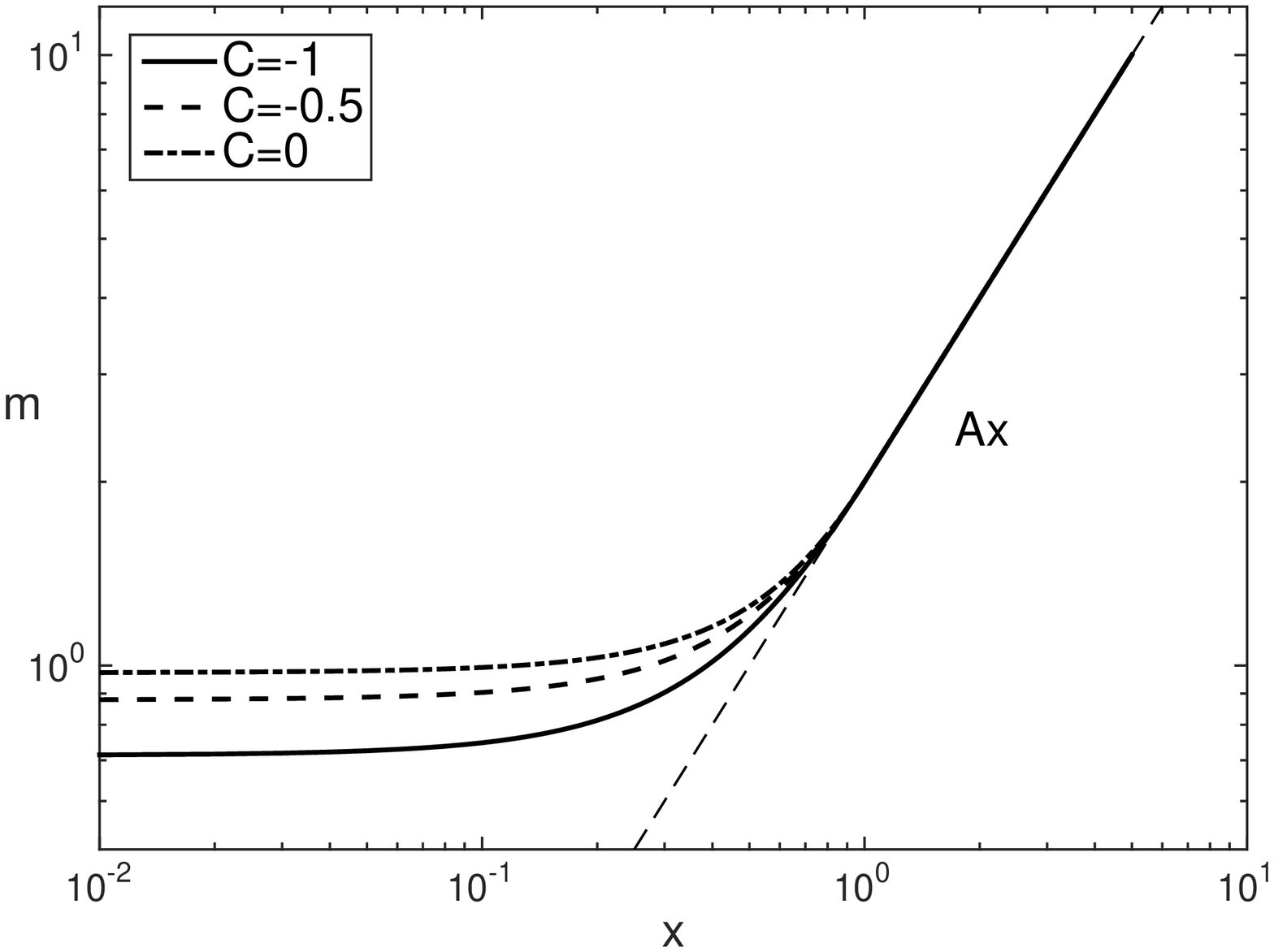}\label{fig: svmas}}
\subfigure[$A=3$]{
   \includegraphics[width=8.5cm]{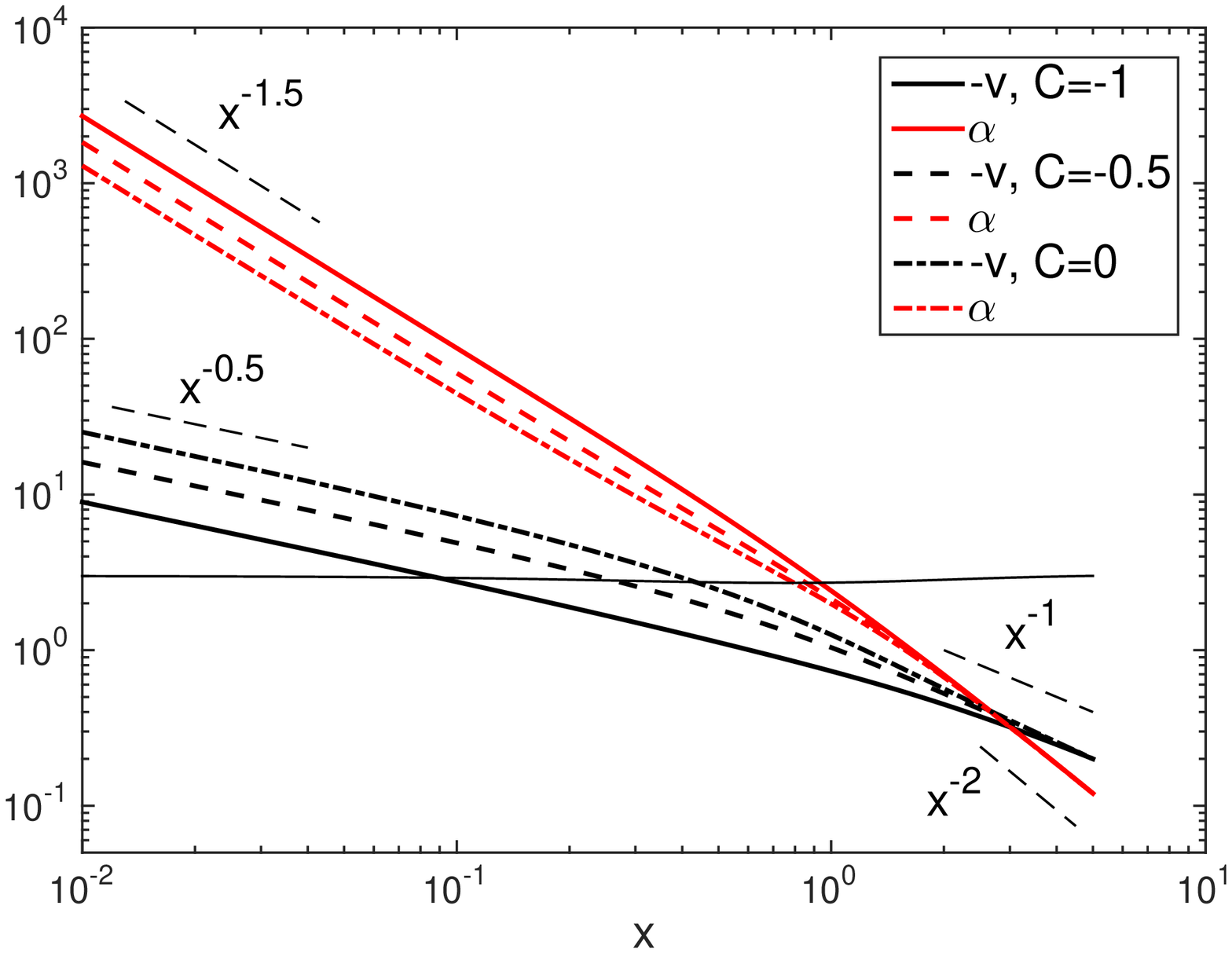}\label{fig: svdc}}
\subfigure[$A=3$]{
   \includegraphics[width=8.5cm]{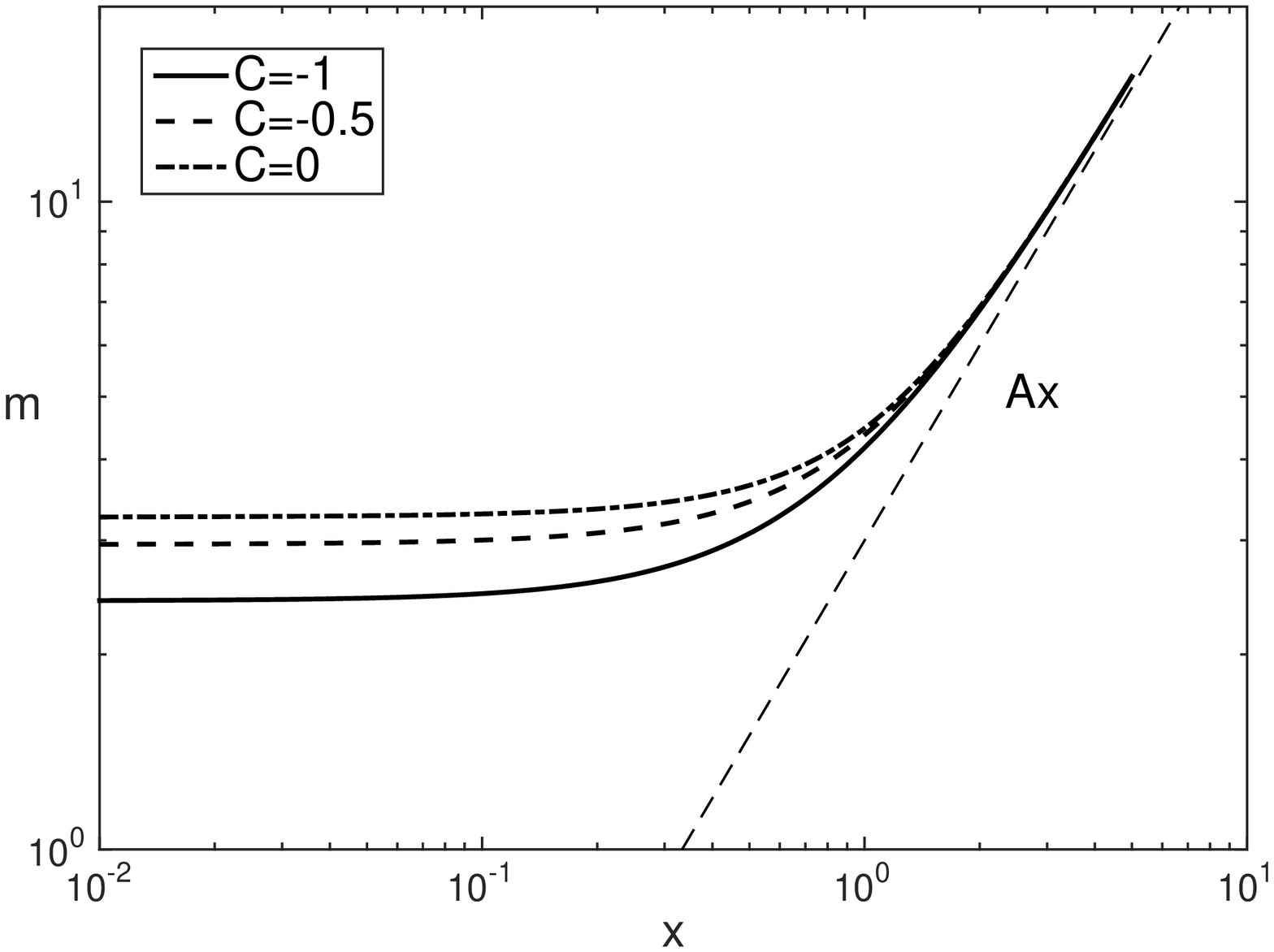}\label{fig: svdcm}}
\caption{Similarity solutions for $-v$, $\alpha$ ((a) and (c)), and $m$ ((b) and (d)) in Case (1)
with the initial condition given by Eq. \eqref{eq: afssvi}. 
The thin solid line in (a) and (c) indicates $ \mathcal{R}$ (Eq. \eqref{eq: genratr}) with $C=-1$.
The short dashed lines indicate the analytically derived asymptotic scalings. }
\label{fig: sv}
\end{figure*}

In Case (1) with an initially small infall velocity, 
the numerical solution to Eqs. \eqref{eq:cp1} and \eqref{eq:cp2} in the entire range of $x$ is presented in Fig.~\ref{fig: sv}.
Our analytical scalings well describe its asymptotic behaviors. 
From the envelope to the inner region, 
$\mathcal{R}$ changes from (Eqs. \eqref{eq: ratthe} and \eqref{eq: fslvden})
\begin{equation}
   \mathcal{R}_\text{the} 
    \sim \alpha x^2   = A 
\end{equation}
to (Eqs. \eqref{eq: ratur} and \eqref{eq: coffs})
\begin{equation}\label{eq: rturffa}
  \mathcal{R}_\text{tur} 
  \sim -\frac{   \alpha x }{  C^2  v} = \frac{   1+ 5 C^2 }{     2 C^2 }.
\end{equation}
Due to the turbulent support, $\mathcal{R}$ remains constant,
which would otherwise increase as $x \rightarrow 0$ in a free-fall regime 
\citep{Sh77}. 
Comparing the cases with different values of $C$, we see that turbulence in the inner region results in 
a lower infall velocity, a higher $\rho$, and a lower constant rate of mass accretion onto a central mass point
\begin{equation}
  \dot{M} = \frac{\partial M(0,t)}{\partial t} =    \frac{a^3 m(0)}{G} 
\end{equation}
compared to the free-fall case.

\begin{figure*}[htbp]
\centering   
\subfigure[]{
   \includegraphics[width=8.5cm]{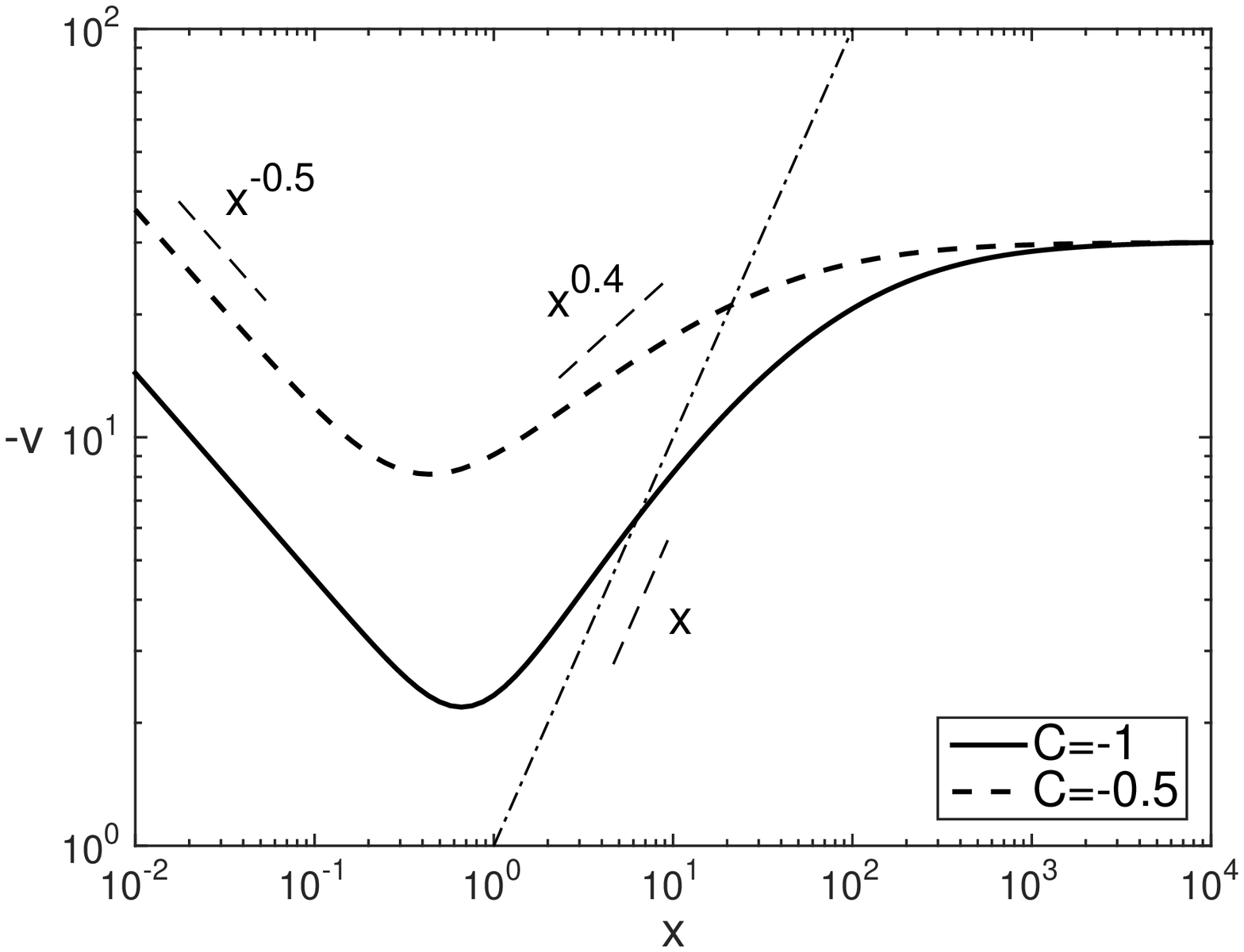}\label{fig: lxv}}
\subfigure[]{
   \includegraphics[width=8.5cm]{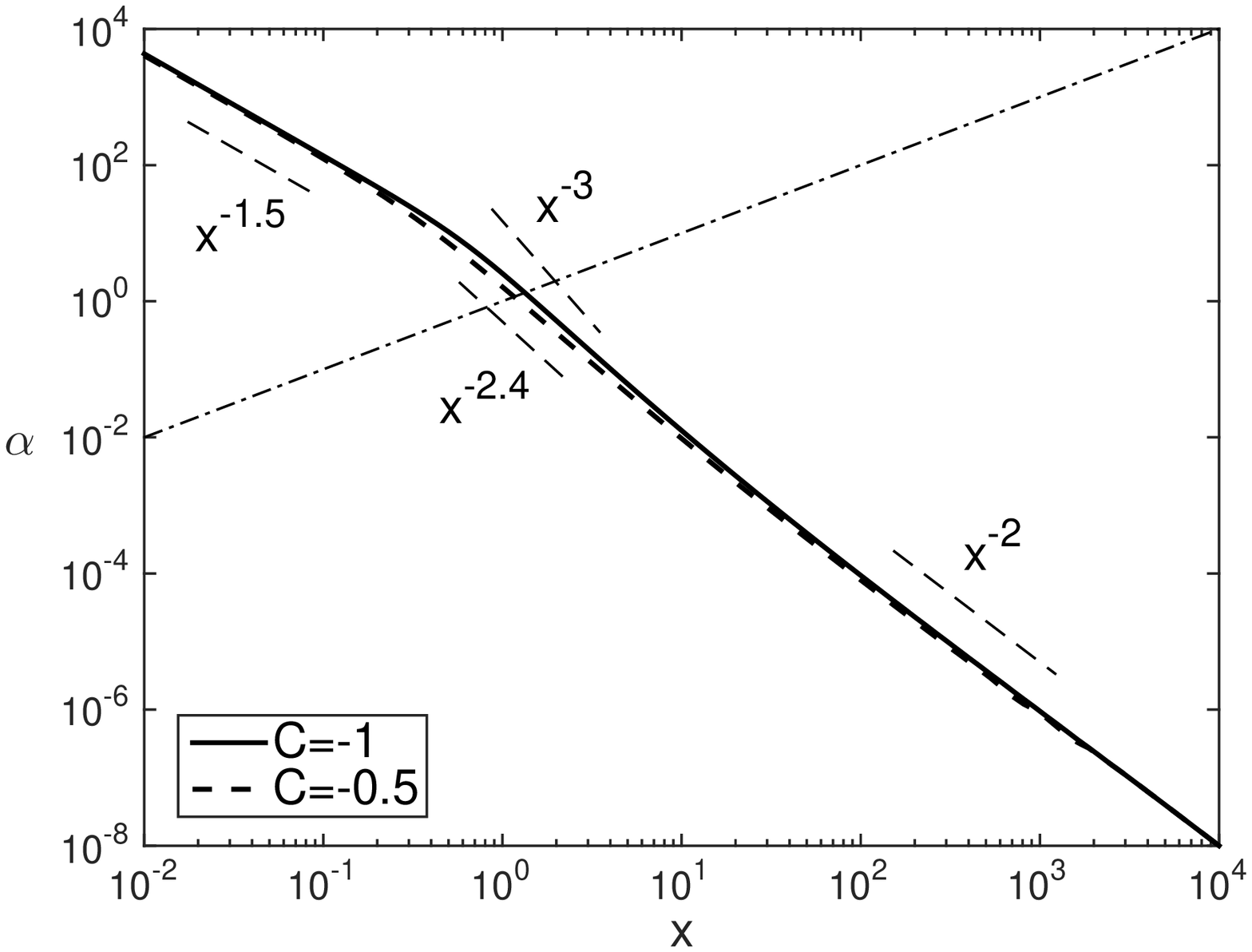}\label{fig: lxd}}
\subfigure[]{
   \includegraphics[width=8.5cm]{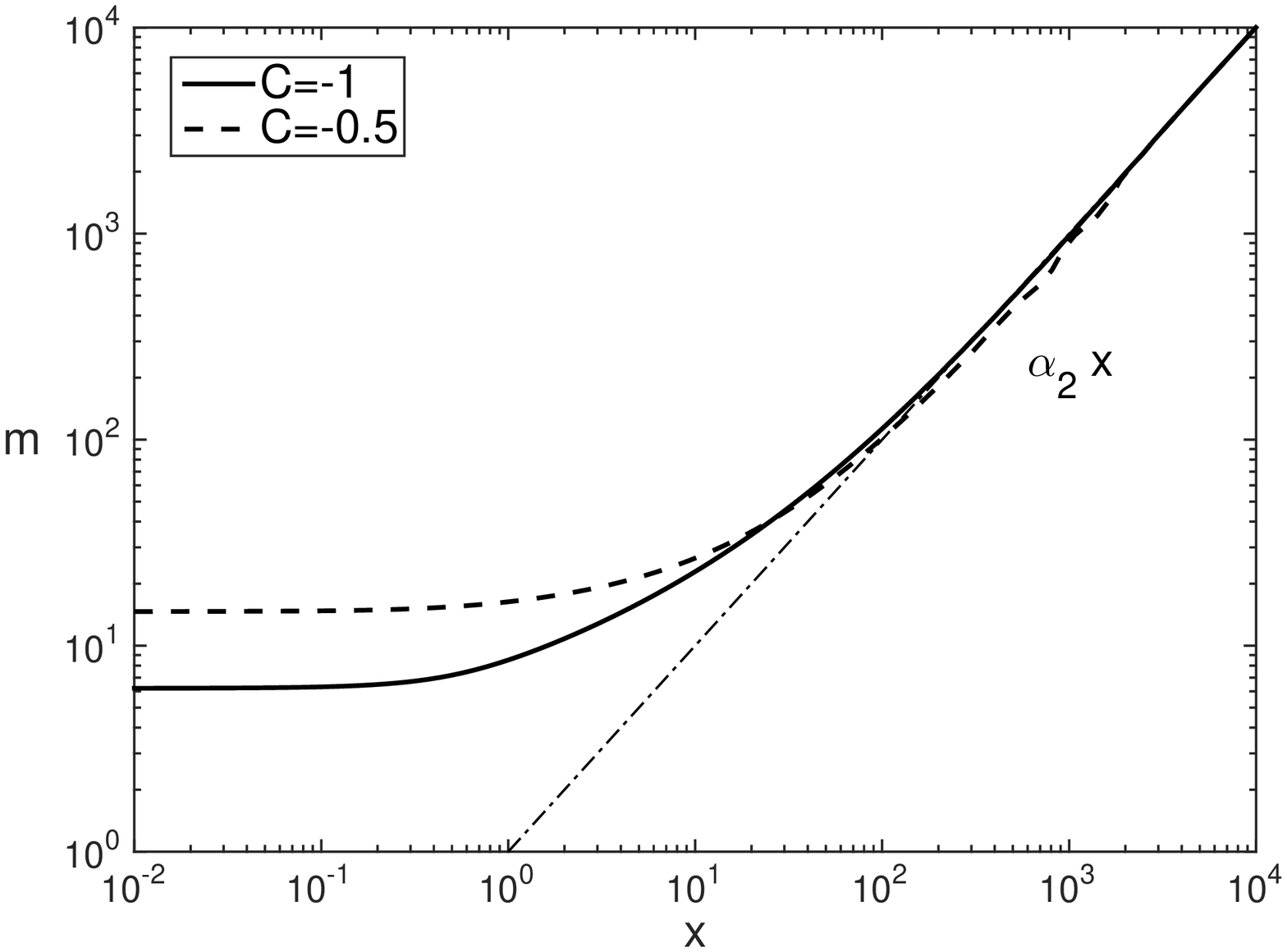}\label{fig: lxm}}
\caption{ Similarity solutions in Case (2) with the initial condition given by Eq. \eqref{eq: csolv}, where $v_2=30$ and $a_2=1$.
The dash-dotted line denotes $|v| = x$ in (a) and (b). }
\label{fig: lx}
\end{figure*}

\begin{figure*}[htbp]
\centering   
\subfigure[]{
   \includegraphics[width=8.5cm]{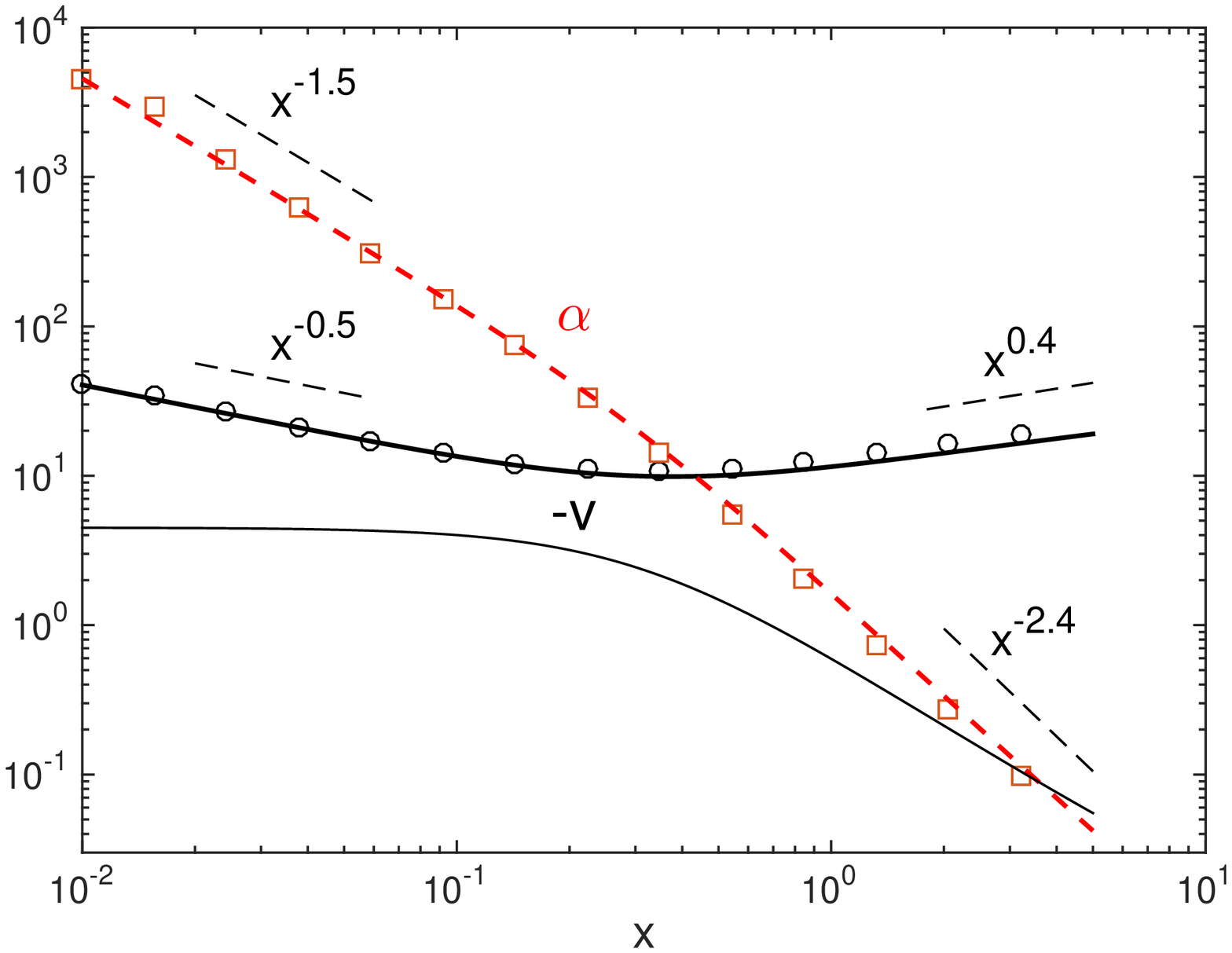}\label{fig: lvc05}}
\subfigure[]{
   \includegraphics[width=8.5cm]{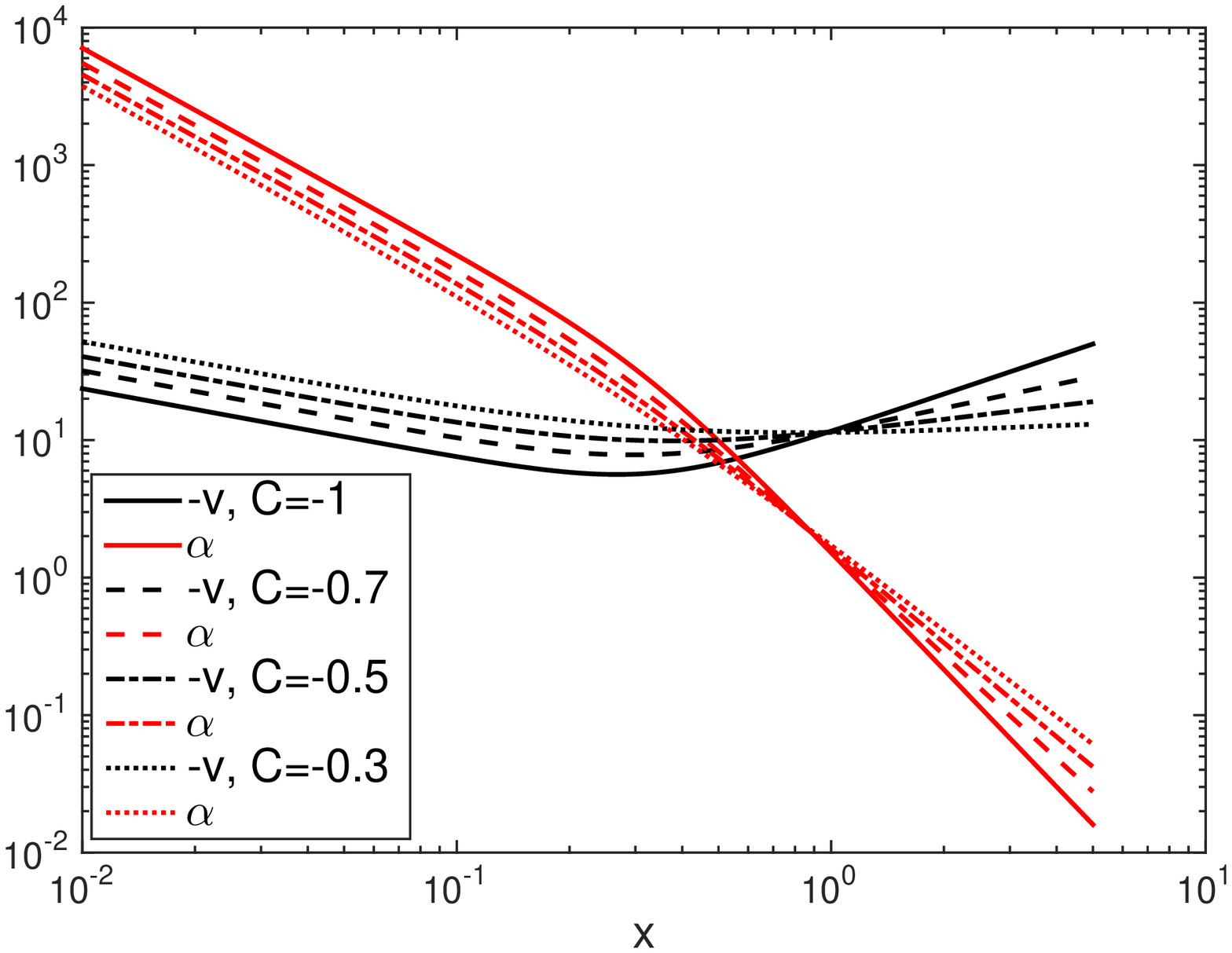}\label{fig: lvdc}}
\subfigure[]{
   \includegraphics[width=8.5cm]{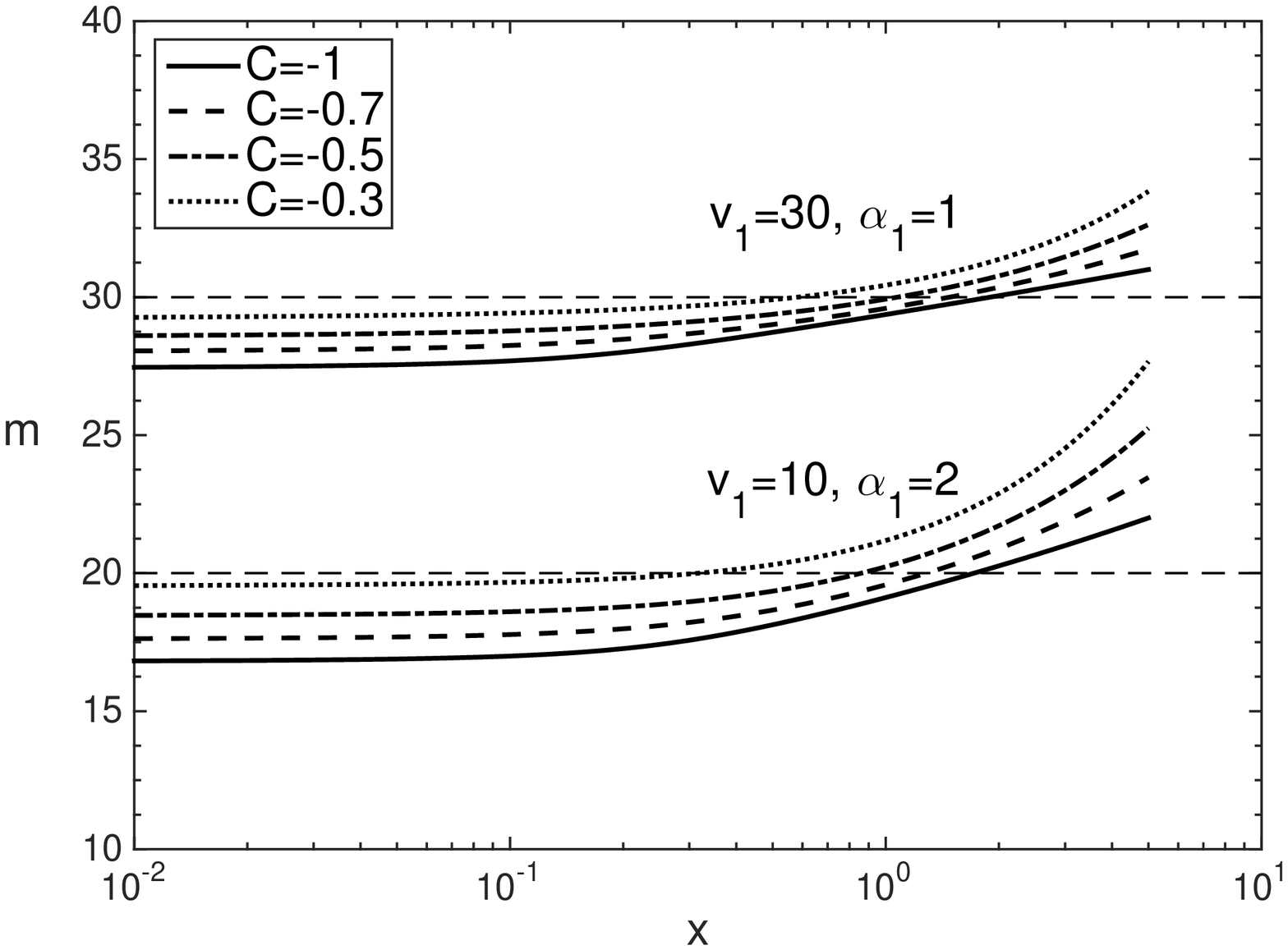}\label{fig: lvmdc}}
\caption{ Similarity solutions with the boundary condition given by Eq. \eqref{eq: decasol}. $v_1 = 10$ and $\alpha_1 = 2$ are adopted 
for (a) and (b). 
In (a), $C=-0.5$, and the thin solid line indicates $\mathcal{R}$ at $C=-0.5$. 
Circles represent the solution derived with the boundary condition given 
by Eq. \eqref{eq: coffs} at a small $x$.  
The thin dashed lines in (c) denote the asymptotic value of $m$ given by Eq. \eqref{eq: asmbtp}. }
\label{fig: lv}
\end{figure*}

In contrast, in Case (2) with an initially large infall velocity, the numerical result 
in the entire range of $x$ is displayed in
Fig.~\ref{fig: lx}.
We see three different regimes with 
(i) constant infall,
(ii) decelerated infall, 
and (iii) accelerated infall. 
To better illustrate the asymptotic scalings in the dynamically unstable regimes (ii) and (iii), 
Fig. \ref{fig: lv} presents the solutions for $x < |v|$ with the boundary condition given by Eq. \eqref{eq: decasol}. 
From regime (ii) to regime (iii), $\mathcal{R}$ changes from (Eqs. \eqref{eq: ratur} and \eqref{eq: decasol})
\begin{equation}
  \mathcal{R}_\text{tur} 
  \sim - \frac{   \alpha x }{  C^2  v} =  \frac{\alpha_1}{C^2 v_1} x^{-\frac{1+5 C^2}{1+C^2}} ,
\end{equation}
which increases with decreasing $x$,
to $\mathcal{R}_\text{tur}$ expressed in Eq. \eqref{eq: rturffa}. 
The change of $\mathcal{R}_\text{tur}$ clearly indicates the transition from turbulent pressure- to 
gravity-dominated dynamics. 
Besides, from Figs. \ref{fig: lxm} and \ref{fig: lvmdc}, we find that the change of $m$ in regimes (ii) and (iii)
is insignificant and it approaches constant toward a small $x$, leading to 
a flat radial profile of $M$ and 
a constant $\dot{M}$ as in Case (1).

\begin{figure*}[htbp]
\centering   
\subfigure[Case (1), C=0]{
   \includegraphics[width=8.5cm]{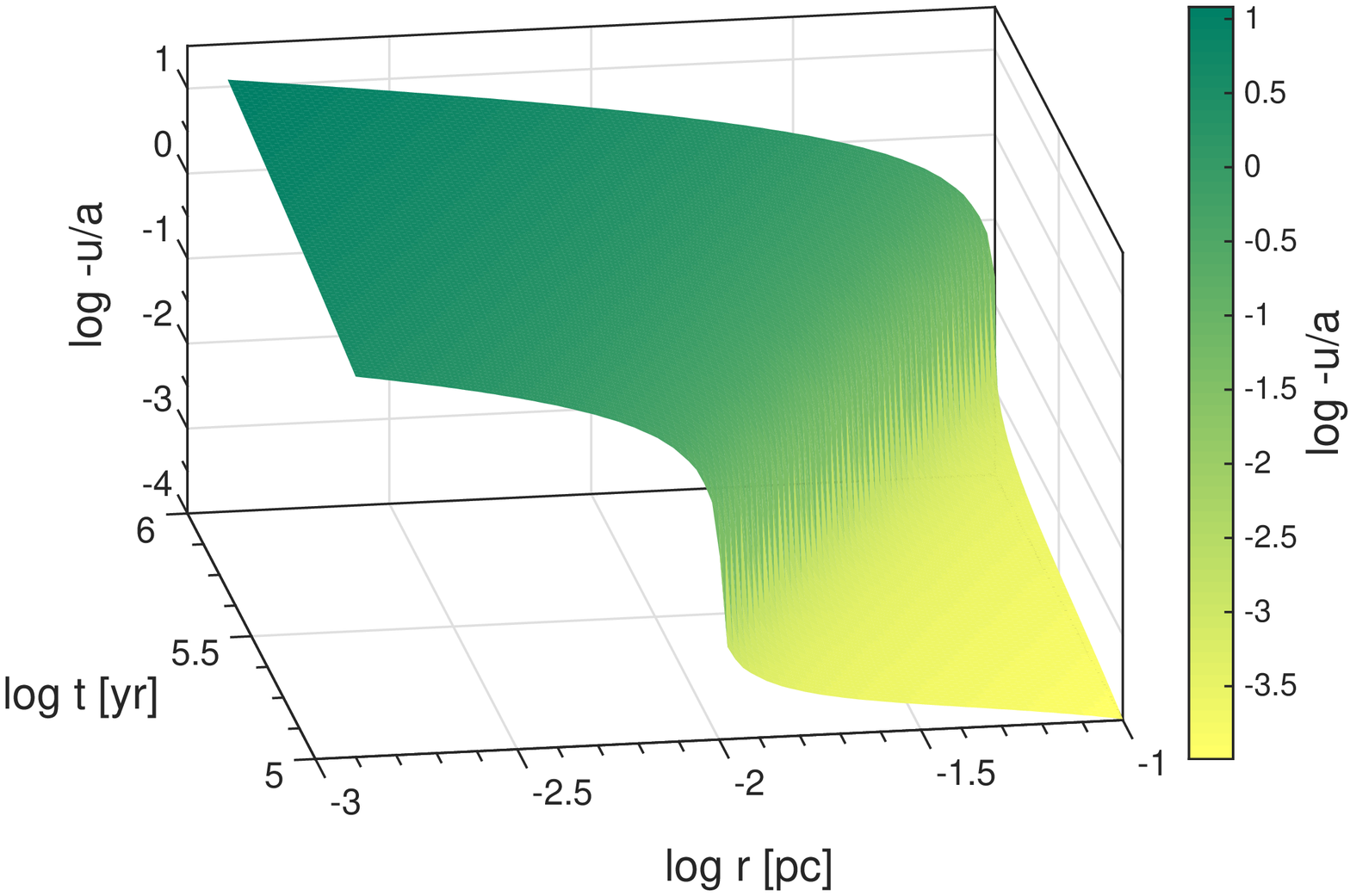}\label{fig.fsvel}}
\subfigure[Case (2), C=-1]{
   \includegraphics[width=8.5cm]{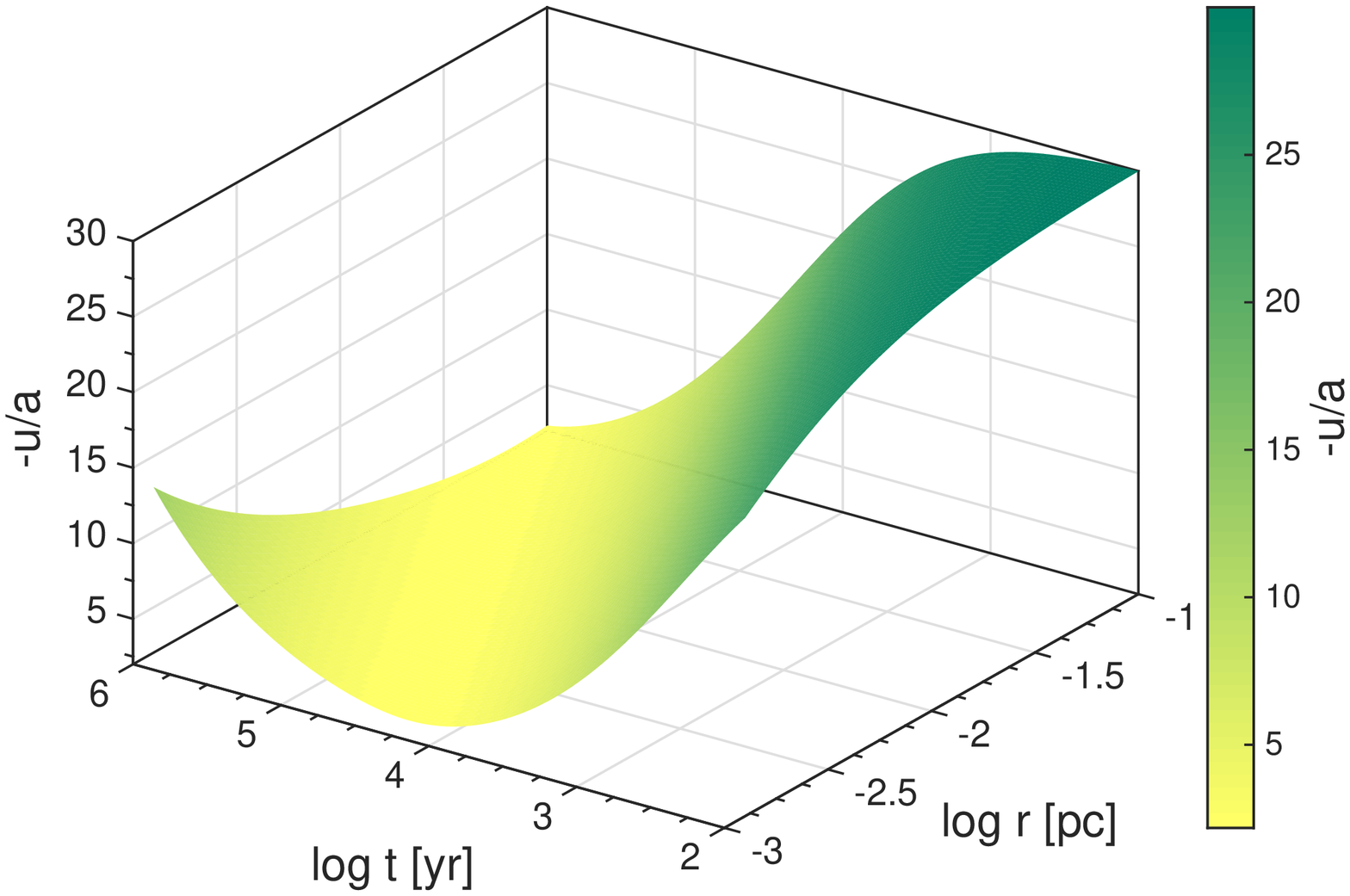}\label{fig: turvel}} 
\subfigure[Case (1), C=0]{
   \includegraphics[width=8.5cm]{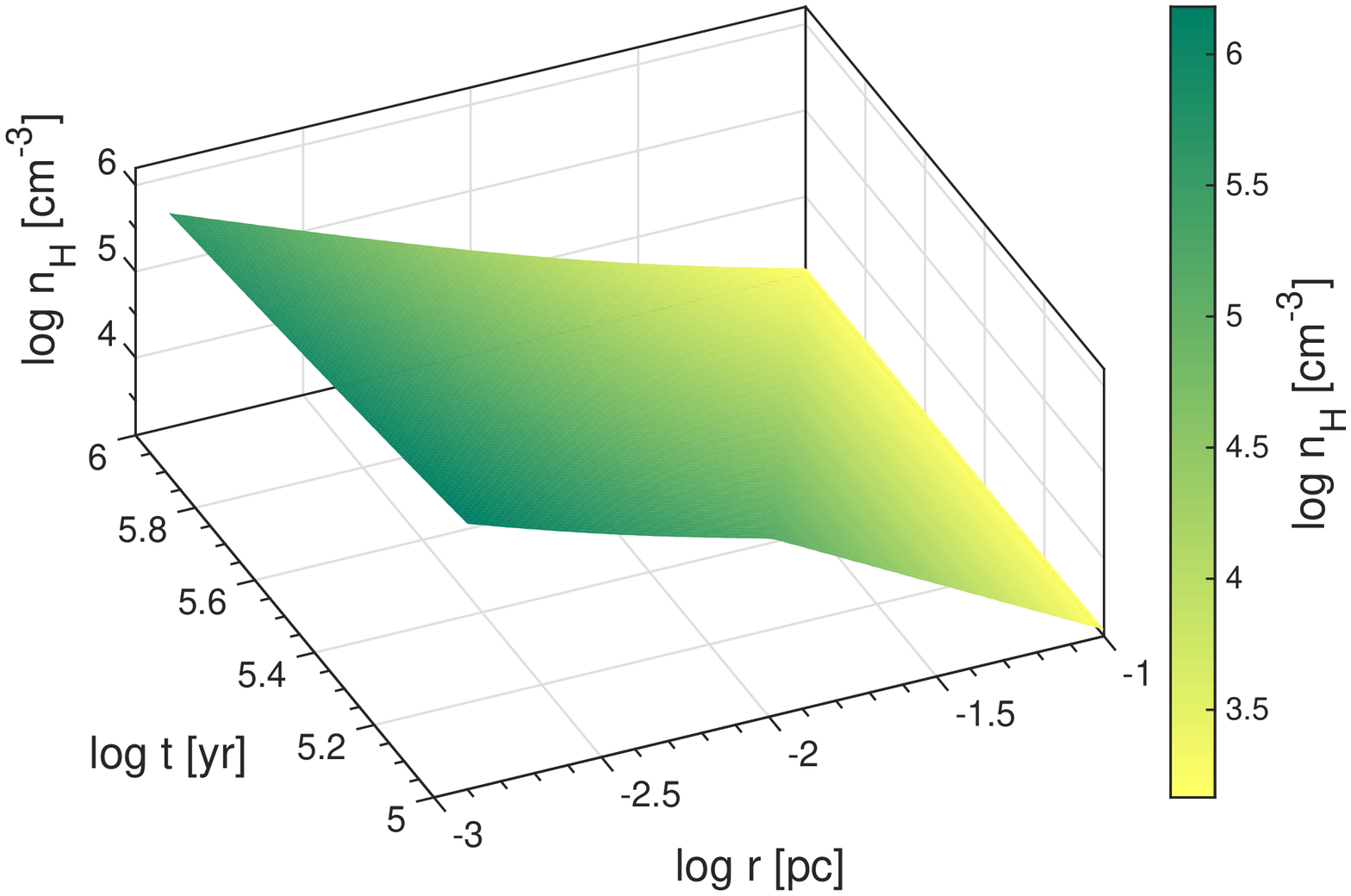}\label{fig: fsden}}
\subfigure[Case (2), C=-1]{
   \includegraphics[width=8.5cm]{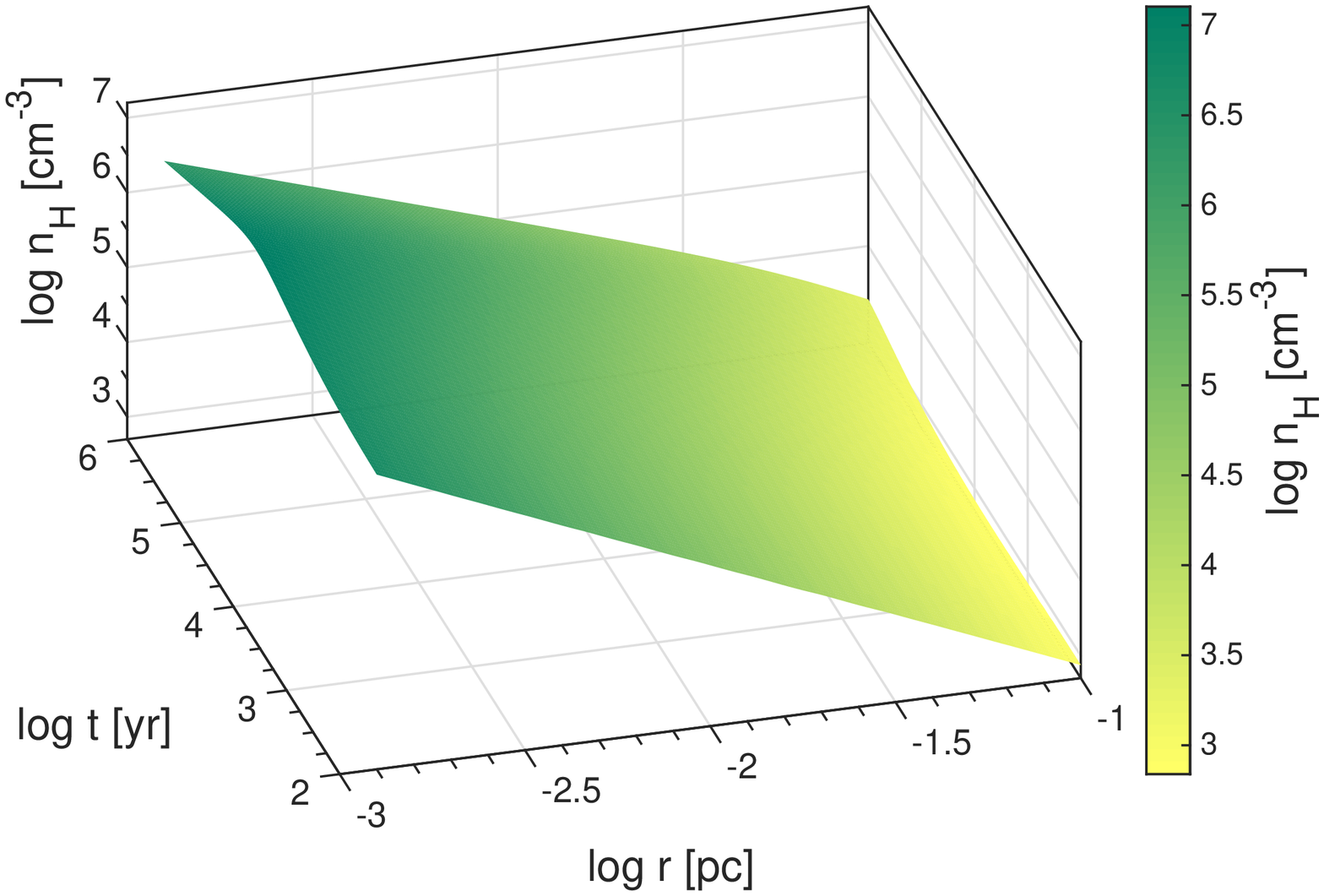}\label{fig: turden}}
\subfigure[Case (1), C=0]{
   \includegraphics[width=8.5cm]{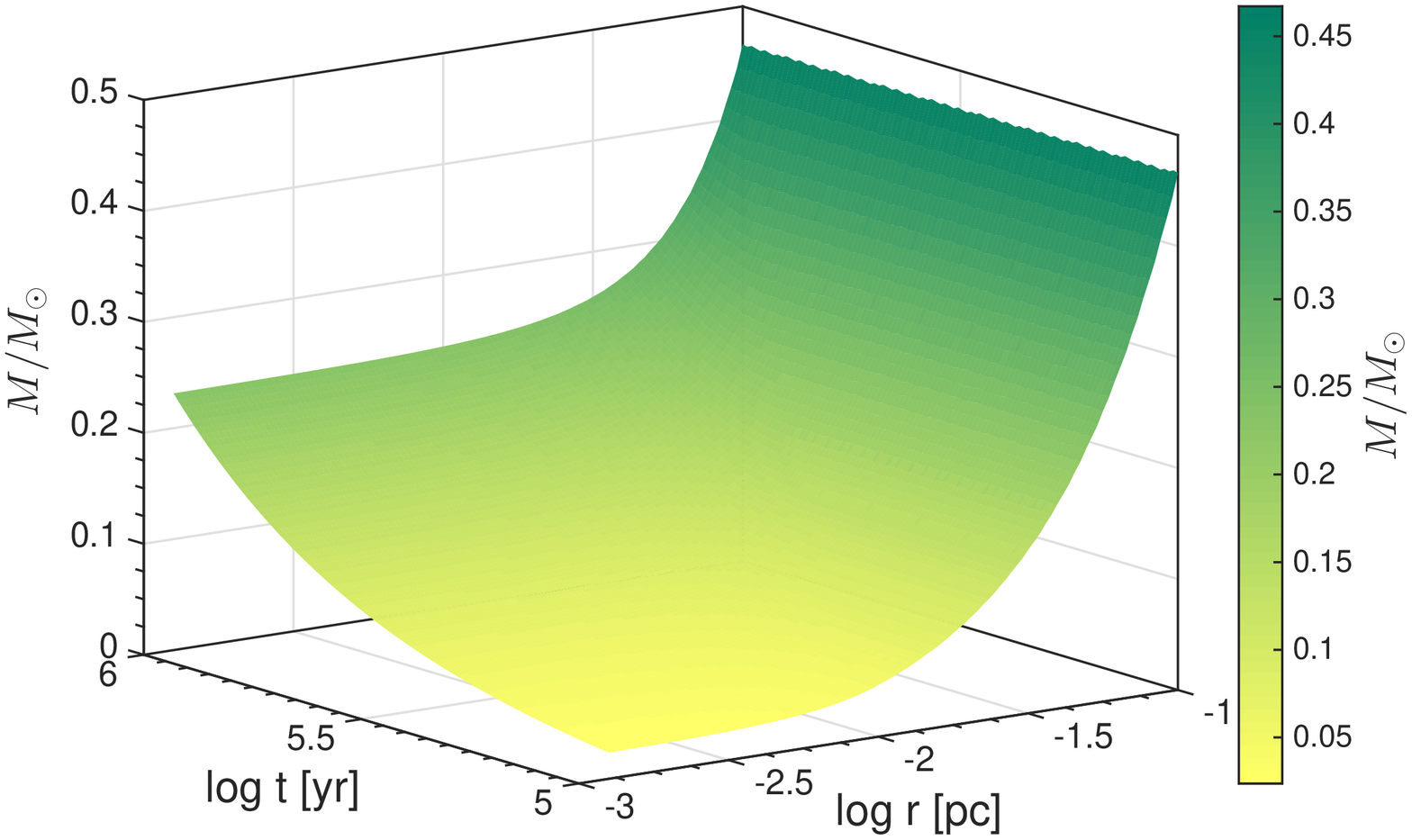}\label{fig: fsmas}}
\subfigure[Case (2), C=-1]{
   \includegraphics[width=8.5cm]{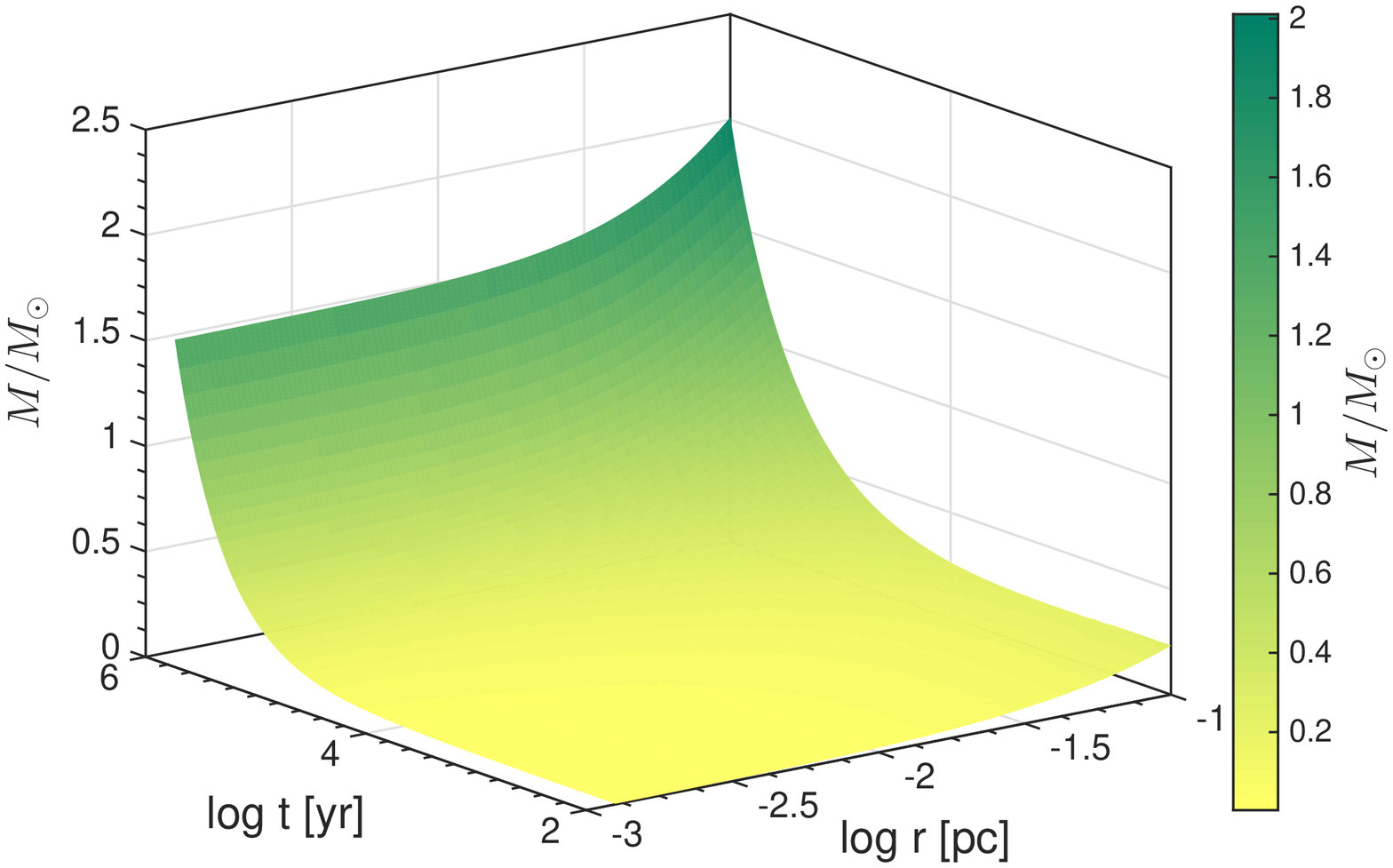}\label{fig: turmas}}
\caption{Time evolution of the radial profiles of $-u(r,t)$ (normalized by $a$), $n_H(r,t)$, and $M(r,t)$ (normalized by $M_\odot$)
in Case (1) with $C=0$ and $A=2.001$
((a), (c), and (e)) and in Case (2) with $C=-1$, $v_2 = 30$, and $\alpha_2 = 1$((b), (d), and (f)). 
}
\label{fig: 3d}
\end{figure*}

The above results clearly demonstrate the importance of gravity-driven turbulence in affecting the collapse dynamics. 
Compared with the isothermal collapse with only thermal pressure, 
turbulence provides additional pressure support against the self-gravity and enables deceleration of the infall. 
Compared with the adiabatic collapse where the released gravitational energy is absorbed, 
due to the dissipation of turbulent energy, the gravity-driven turbulence is incapable of halting the gravitational contraction.
At a sufficiently small $x$, the solution has the free-fall scaling.
Fig. \ref{fig: 3d} shows the numerically solved $-u(r,t)$, the number density of atomic hydrogen
$n_H (r,t) = \rho(r,t)/m_H$, where $m_H$ is the mass of hydrogen atom, 
and $M(r,t)$. 
As a comparison, we present both cases of a non-turbulent collapse with a quasi-static envelope 
(Figs. \ref{fig.fsvel}, \ref{fig: fsden}, and \ref{fig: fsmas}, corresponding to Fig. \ref{fig: svda} with $C=0$)
and a turbulent collapse with an initially large infall velocity (Figs. \ref{fig: turvel}, \ref{fig: turden}, and \ref{fig: turmas}, 
corresponding to Figs. \ref{fig: lxv} and \ref{fig: lxd} with $C=-1$). 
Note that we adopt the values of parameters here only for illustrative purposes, but not for detailed comparisons with specific observations. 
In the former case, 
we can easily see that the outward moving expansion wavefront separates the free-fall regime and the quasi-static regime.
While in the latter case the collapse exhibits a more complex behavior. 
Comparing the two scenarios, 
despite the different initial conditions and different levels of turbulence, the same scalings of velocity, density, and mass apply to the central region
after a sufficiently long time, showing the dominance of self-gravity at the center.

\section{Discussion}

In the presence of turbulence, the stochastic wandering of magnetic field lines naturally takes place as a result of 
turbulent energy cascade and turbulent mixing of magnetic fields. 
Consequently, 
the reconnection of wandering magnetic fields are much more efficient than the microscopic magnetic reconnection
\citep{LV99}. 
The latter relies on the resistive diffusion in a conducting fluid or the ambipolar diffusion in a partially ionized medium. 
On length scales where turbulence exists, it is the turbulent diffusion of magnetic fields that dominates over the above microscopic diffusion 
processes.
The turbulent reconnection of magnetic fields violates flux freezing and allows turbulent diffusion of magnetic fields, 
which has been termed ``reconnection diffusion (RD)"
\citep{Laz05}.
The diffusion rate only depends on the turbulence properties
\citep{KL09}.

\begin{figure}[!h]
\centering   
   \includegraphics[width=8.5cm]{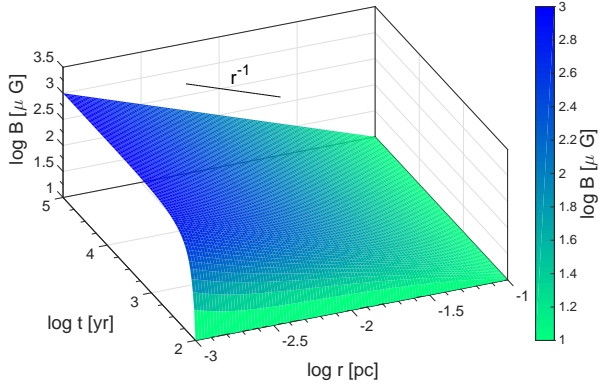}
\caption{Time evolution of magnetic field profile as the numerical solution to Eq. \eqref{eq: odine}
in the decelerated infall regime in Case (2), with 
$v_1 = 10$ (Eq. \eqref{eq: vahn}) and 
$C=-1$. 
The initial 
uniform magnetic field has a strength $10~ \mu$G.}
\label{fig: magte}
\end{figure}

To illustrate the effect of RD induced by gravity-driven turbulence,
in Fig. \ref{fig: magte}, we present the evolution of magnetic field profile in the decelerated infall regime in Case (2)
(see Appendix \ref{app:rd} for the detailed calculations). 
We see that 
the RD rapidly balances the gravitational drag and stabilizes the magnetic field profile to have the form consistent with Eq. \eqref{eq: sstemf},
\begin{equation}\label{eq: stbrpf}
     B_s (r) = 10~ \mu \text{G} \Big(\frac{r}{0.1 ~\text{pc}}\Big)^{-1}.
\end{equation}
It suggests that the RD results in an efficient expulsion of magnetic fields 
and prevents the accumulation of magnetic flux in a collapsing region.

The application of RD to star formation processes 
\citep{Sant10,LEC12,Le13,Laz14,LiM15,Moc17}
demonstrates that RD 
leads to violation of flux-freezing 
\citep{Eyin13}
and is indispensable for solving
the ``magnetic flux problem"
\citep{Mes56},
accounting for the observed supercritical molecular clouds and cores 
\citep{Crut10}
and the observed strengths of surface magnetic fields of stars 
\citep{Joh04}. 
RD also mitigates the magnetic braking ``catastrophe" and allows the formation of centrifugally supported circumstellar disks
(\cite{San12,San13,Die16}; see also \cite{Gra18}).
Most earlier studies on RD involved externally driven turbulence, 
e.g., the interstellar turbulence driven by supernova explosions on large scales. 
\footnote{In most MHD simulations of the interstellar turbulence, turbulence is continuously forced at a large driving scale
to simulate the externally driven turbulence for a system on scales smaller than the driving scale of turbulence. 
In the simulations with decaying turbulence, additional turbulence can be internally driven after the gravitational contraction 
of the system initiates.}
Differently,  
here we find that RD naturally occurs during the gravitational collapse without an external source for driving turbulence.

In a weakly ionized and magnetized core, besides RD,
the ambipolar diffusion (AD) of magnetic fields due to ion-neutral drift also takes place. 
The comparison between the rates of AD and RD shows 
\begin{equation}
\begin{aligned}
    \frac{ \omega_d }{v_t/r} &= \frac{\xi_n V_A^2 }{6 \nu_{ni}   v_t r} 
    = \frac{\xi_n v_t }{6 \nu_{ni}    r}   \\
 &   = 3.2\times10^{-4} \Big(\frac{v_t}{0.1 \text{km s}^{-1}}\Big) \Big(\frac{r}{0.1 \text{pc}}\Big)^{-1} \\
  &   ~~~~   \Big(\frac{n_H}{10^4 \text{cm}^{-3}}\Big)^{-1} \Big(\frac{n_e/n_H}{10^{-6}}\Big)^{-1}, 
\end{aligned}
\end{equation}
where $\xi_n = \rho_n /\rho$ is the neutral fraction with the neutral mass density $\rho_n$ and the total mass density $\rho$, 
$\nu_{ni} = \gamma_d \rho_i$ is the neutral-ion collision frequency with the drag coefficient $\gamma_d=3.5\times10^{13}$cm$^3$g$^{-1}$s$^{-1}$
\citep{Drai83,Shu92} 
and the ion mass density $\rho_i$, and
$n_H$ and $n_e$ are 
number densities of the atomic hydrogen and electrons. 
Here we also assume $v_t = V_A$, where $V_A$ is the Alfv\'{e}n speed,
and $m_i=29 m_\text{H}$, $m_n=2.3 m_\text{H}$ 
as the mean molecular mass of ions and neutrals in a core
\citep{Shu92},
where $m_\text{H}$ is the hydrogen atomic mass. 
Evidently, in the presence of turbulence, AD is subdominant compared to RD.

\section{Conclusions}

The effect of turbulence on gas dynamics varies with the length scale of interest. 
For the interstellar turbulence with a driving scale $\sim 50-100$ pc, 
shear Alfv\'{e}nic motions and compressive motions in supersonic turbulence play an important role in 
shaping the density structures within the inertial range of the interstellar turbulence 
\citep{Pad01,Fed10,XuZ16,XuZ17,RoG18,Moc18}.
For the gravity-driven turbulence in a contracting core considered here, the driving scale is small, and thus 
the internal turbulent motions provide pressure support for the surrounding density shells.
We found that the gravity-driven turbulence can slow down the
gravitational infall and 
mass accretion.

Compared with the Kolmogorov scaling $v_t \propto r^{1/3}$ 
(e.g., \cite{Qi18}) 
or the Larson's scaling $v_t \propto r^{1/2}$
\citep{Lars81,Mye83}
in the inertial range of externally driven turbulence,
we found that the gravity-driven turbulence can give rise to various velocity dispersion profiles,
$v_t\propto r^\alpha$ with $0\leq \alpha \leq 1$ in the outer region of a dynamically contracting core at an early time of its evolution, 
and $v_t \propto r^{-1/2}$ toward the center in a quasi-statically contracting core or a dynamically contracting core at a late time of its evolution 
(Fig. \ref{fig: 3d}).
Our analytical scalings are consistent with earlier numerical results in the parameter space of the simulations 
\citep{RobG12}.
Observations suggest that 
the non-thermal line width-size relation 
of massive cores is flatter than that of low-mass cores 
\citep{Cas95}.
\citet{Plu97}
found no statistically significant line width-size relationship 
or a positive correlation between line width and density for very massive cores. 
These unexpected observational findings support the theoretical picture of gravity-driven turbulence in molecular cloud cores 
(see also \cite{Mu15}). 
More detailed and quantitative comparisons will be carried out in our future work.

As an important implication of the current study, 
the gravity-driven turbulence not only influences the dynamics of a collapsing core, but also enables 
an efficient diffusion of magnetic fields. 
At the balance between the gravitational drag and diffusion, 
a stationary radial profile of magnetic field can be reached,
with the slope depending on the  
fraction of gravitational potential energy converted to the turbulent kinetic energy. 
\\
\\


S.X. acknowledges the support for Program number HST-HF2-51400.001-A provided by NASA through a grant from the Space Telescope Science Institute, which is operated by the Association of Universities for Research in Astronomy, Incorporated, under NASA contract NAS5-26555.
A.L. acknowledges the support from grant
NSF DMS 1622353.

\appendix 

\section{RD due to gravity-driven turbulence}\label{app:rd}

Here we consider gravitational contraction of weak magnetic fields as suggested by observations 
\citep{Crut10}. 
Since the gravitational drag only occurs along the radial direction,
we adopt the 
one-dimensional induction equation, i.e., the magnetic convection-diffusion equation 
\citep{Luh84},
\begin{equation}\label{eq: odine}
    \frac{\partial B}{\partial t} =- \frac{\partial}{\partial r} (B u) +  \frac{\partial}{\partial r} \Big(\kappa \frac{\partial B}{\partial r}\Big),
\end{equation}
to discuss the RD of dynamically insignificant magnetic fields.
The first term on the RHS of Eq. \eqref{eq: odine} describes the magnetic fields being dragged inward with the infalling flow, 
which alone corresponds to flux freezing. 
The second term describes the RD of magnetic fields, where the diffusion coefficient is 
\begin{equation}
     \kappa = v_t r  = C u r
\end{equation}
for super-Alfv\'{e}nic turbulence with the turbulent energy exceeding the magnetic energy. 

Under the effect of RD, when the diffusive term becomes comparable to the convective term, i.e., 
\begin{equation}\label{eq: sstbv}
     Bu = \kappa \frac{\partial B}{\partial r},
\end{equation}
the evolution of magnetic fields reaches a steady state. 
The steady-state magnetic field is 
\begin{equation}\label{eq: sstemf}
     B_s (r) = B_f \Big(\frac{r}{r_f}\Big)^\frac{1}{C}, 
\end{equation}
where $B_f$ is the field strength at a reference radius $r_f$.
We see that 
$B_s(r)$ does not depend on the functional form of $u$ as it is canceled out in Eq. \eqref{eq: sstbv}, 
but only depends on $C$.

\bibliographystyle{apj.bst}
\bibliography{xu}

\end{document}